\documentclass[12pt]{iopart}

\usepackage{graphicx}

\begin{document}

\title[]{A d-Au data-driven prediction of cold nuclear matter effects on $J/\psi$ production in Au-Au collisions at RHIC}

\author{Rapha\"el Granier de Cassagnac}

\address{Laboratoire Leprince-Ringuet, \'Ecole polytechnique/IN2P3, Palaiseau, 91128 France}
\ead{raphael@in2p3.fr}
\begin{abstract}
%
I present here a new Glauber-inspired approach to derive normal cold nuclear matter effects on $J/\psi$ production in Au-Au collisions.
In an as much as possible model-independent way, it extrapolates the centrality dependent yields from d-Au to Au-Au collisions.
I then compare to the new Au-Au measurements shown by the PHENIX experiment in this conference. In the most central collisions, $J/\psi$ survival probabilities beyond cold nuclear matter effects are found to be $44\pm23\%$ at forward ($y=1.7$) and $25\pm12\%$ at mid ($y=0$) rapidities.
\end{abstract}


\section{Cold nuclear matter effects}

$J/\psi$ production is sensitive to nuclear matter and subtracting the normal suppression is a key point in any interpretation of quarkonia production in heavy ion collisions. Data from p-A or d-A collisions provide good insights
into cold nuclear matter effects.

At SPS energies, a single $J/\psi$-nucleon absorption cross-section of $\sigma^{J/\psi}_{abs}=4.18 \pm 0.35$~mb can describe the observed $J/\psi$ suppression in a wide range of p-A systems, as well as in S-U and peripheral Pb-Pb or In-In collisions ~\cite{Alessandro:2004ap,Arnaldi:2006ee}. However, according to various theoretical predictions (such as~\cite{Eskola:1998df}), shadowing (or rather anti-shadowing) could play a role at the SPS regime, namely for momentum fraction $x$ of the order of $10^{-1}$. The SPS experiments can hardly address this issue
because of their limited rapidity (or~$x$) range. However, a large
rapidity asymmetry observed in $J/\psi$ yields in p-A collisions
provides a hint that a nuclear effect is playing a role beyond simple absorption~\cite{Alessandro:2006jt}.

At RHIC energies, 
the PHENIX experiment has measured~\cite{Adler:2005ph} $J/\psi$ nuclear modification factors in d-Au collisions ($R_{dAu}$). It covers a wide rapidity range, from $-2.2$ to 2.4, corresponding to momentum fractions $x$ of gluons
from $10^{-3}$ to $10^{-1}$. It also includes the first ever centrality dependence measurement of $J/\psi$ production in \emph{p-A like} collisions. According to~\cite{Klein:2003dj}, both the rapidity and centrality dependencies can be reproduced in terms of inhomogeneous shadowing
added to a moderate (not larger than 3~mb) normal nuclear absorption. In~\cite{Vogt:2005ia}, the same author derives Au-Au nuclear modification factors ($R_{AuAu}$), assuming various shadowing schemes and absorption cross-sections.
Another attempt to derive a reference from d-Au can be found
in~\cite{Karsch:2005nk} where the authors compute dissociation cross-sections
from the centrality dependence of $R_{dAu}$. They assume that all nuclear
effects are proportional to $\exp(-\rho_0 \sigma_{diss} L)$,
$\rho_0$ being the normal nuclear density and $L$ the average length
of nuclear matter seen by the $J/\psi$. However, nuclear effects such as shadowing do not follow this centrality dependence.


\section{From d-Au to Au-Au at RHIC}

\begin{figure}
\begin{center}
  \begin{tabular}{cc}
  \includegraphics[width=6.8cm]{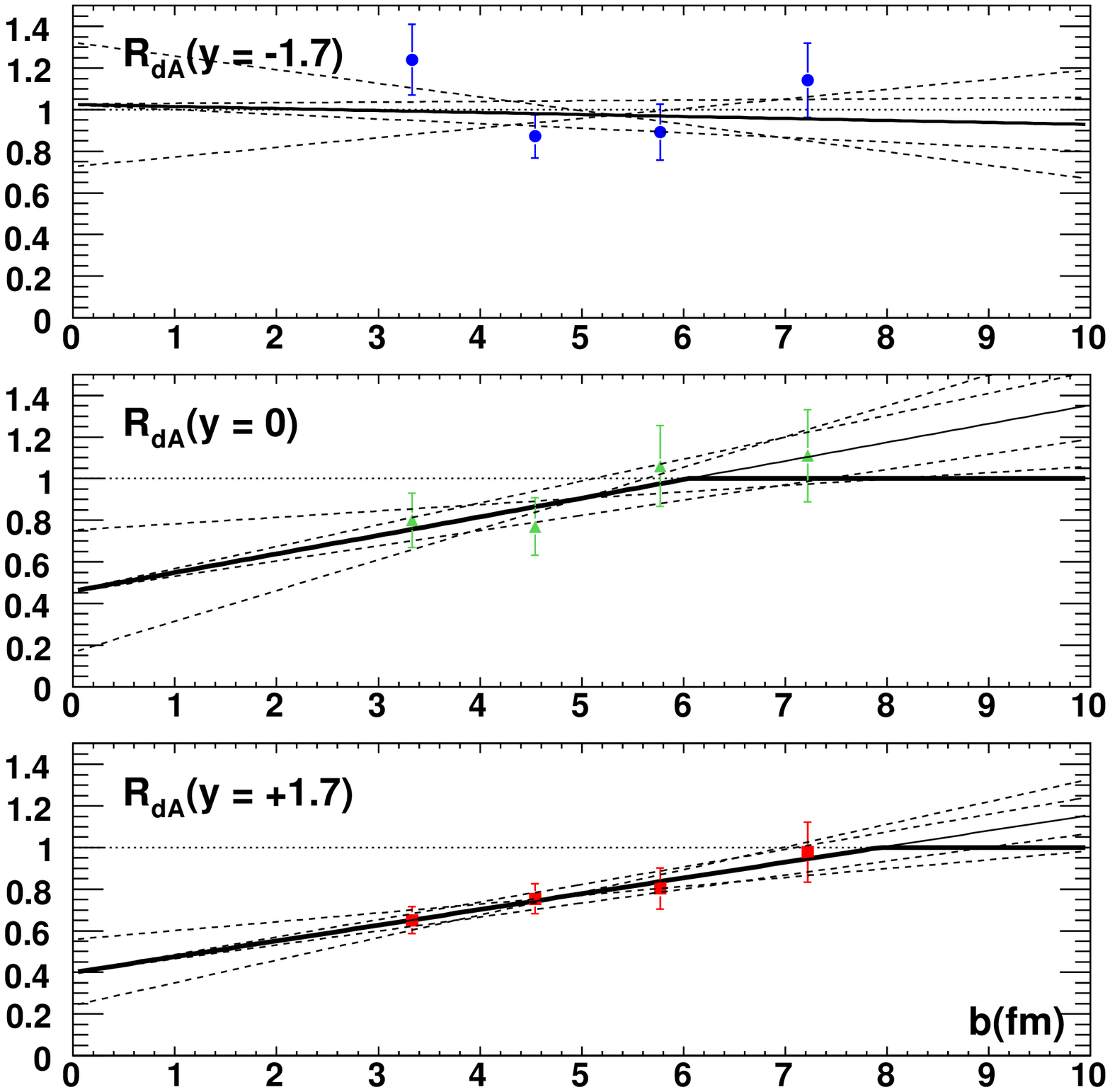} & \includegraphics[width=6.8cm]{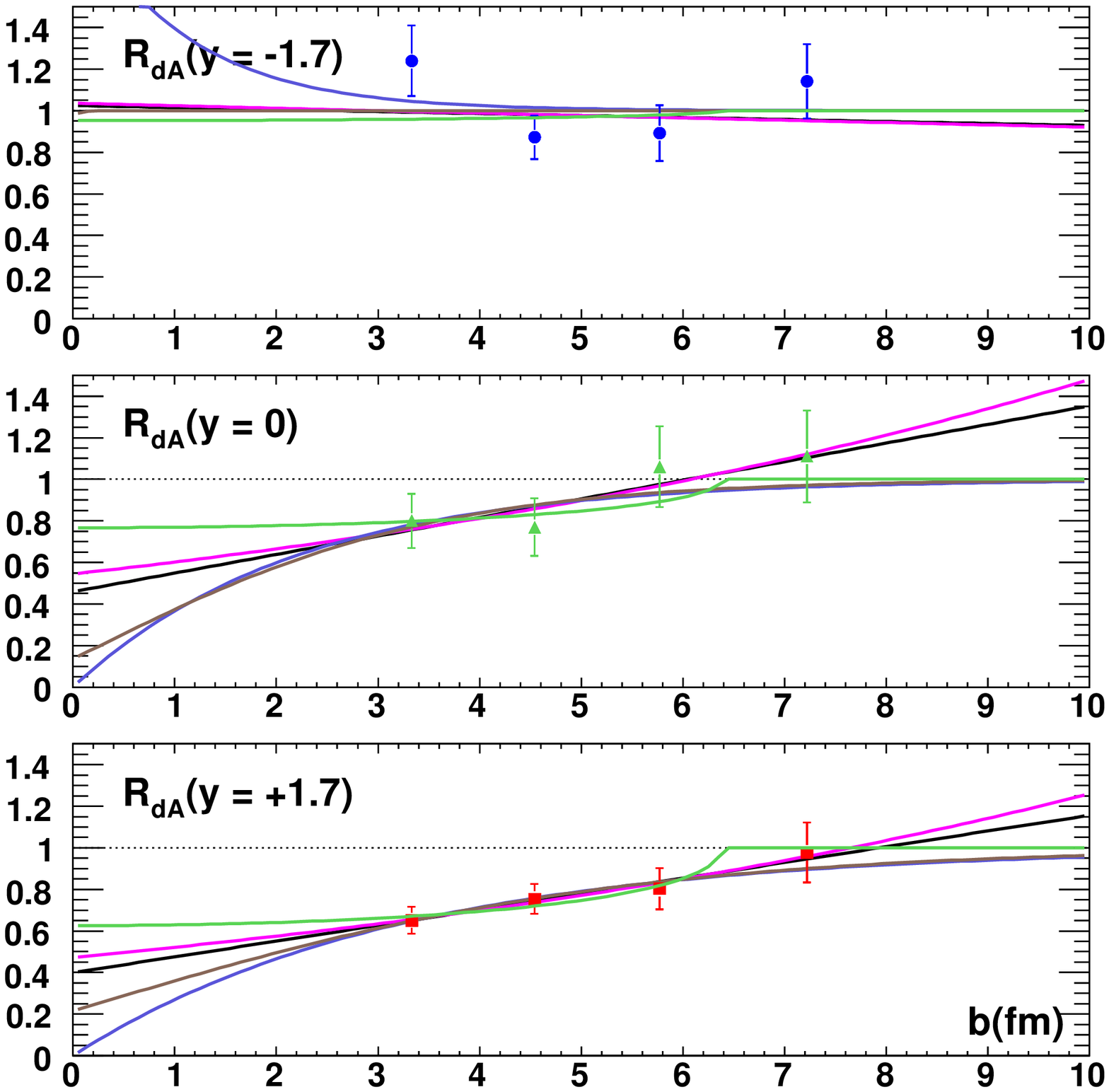}
  \end{tabular}
\end{center}
\vspace{-2ex} \caption{Phenomenological fits to $R_{dAu}$ versus impact parameter. Left) linear fits and one standard deviation variations. Right) other fits for systematic studies: $A \exp(b/B)$, $1-(1-A)\exp(-b/B)$, $A+(1-A) \tanh(b/B)$ or $\exp (-A\sqrt{R^2-b^2})$.}
\label{Fig1}
\vspace{-2ex}
\end{figure}

I propose here an alternate method, with the concerns to be as much data-driven as possible and to derive realistic uncertainties. To do so, I first perform phenomenological fits of the modification factors $R_{dAu}(y,b)$ as a function of the impact parameter $b$ (given by a Glauber model), for the three rapidities $y$ probed by PHENIX~\cite{Adler:2005ph}.
The simplest working fits are linear (figure~\ref{Fig1} left). A cut-off at $R_{dAu}=1$ is applied at $y=0$ and $y=+1.7$ for high impact parameters to reflect the fact that grazing Au-Au collisions are expected to behave like (binary-scaled) p-p collisions. This will prevent $R_{AuAu}$ from reaching unphysically high values (well above unity) at high impact parameters.

I then run a Au-Au Glauber model~\cite{Glauber:1959}
 simulation. For each Au-Au collision occurring at a given impact parameter $b_{AuAu}$, the $N_{coll}$ elementary nucleon-nucleon collisions are randomly distributed (following Woods-Saxon nuclear densities) providing the
locations $b^i_1$ and $b^i_2$ of each collision $i$, relative to the
center of nucleus~1 and nucleus~2. For the considered Au-Au collision,
the predicted nuclear modification factor $R_{AuAu}$ is given by the following
summation over the elementary collisions:
\begin{equation}
 R_{AuAu} (y, b_{AuAu}) = \sum^{N_{coll}}_{i=1} \Big( R_{dAu}(-y,b^i_1) \times R_{dAu}(+y,b^i_2) \Big) / N_{coll}.
 \label{Eq1}
\end{equation}

\begin{figure}
\begin{center}
  \begin{tabular}{cc}
  \includegraphics[width=6.8cm]{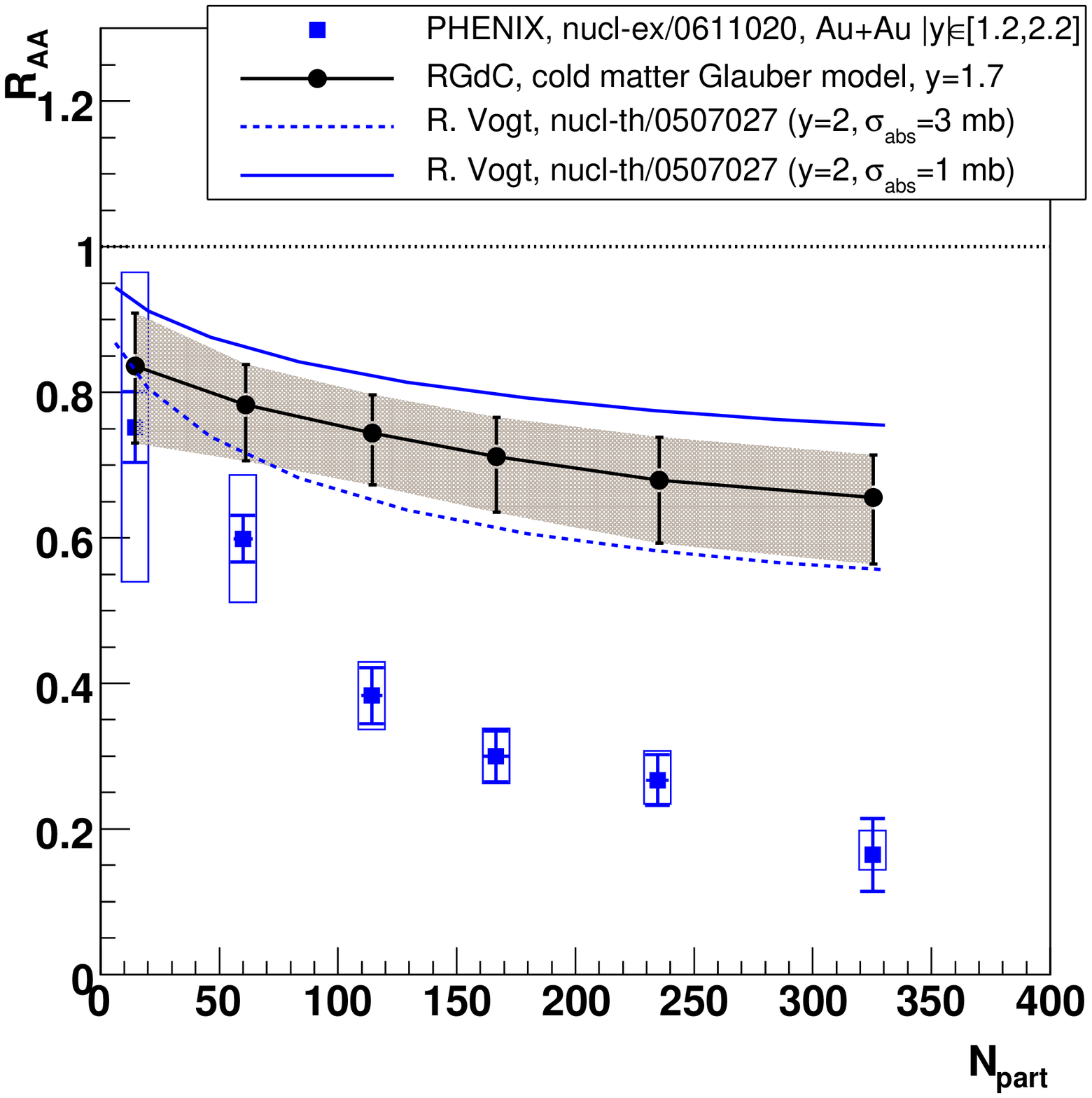} & \includegraphics[width=6.8cm]{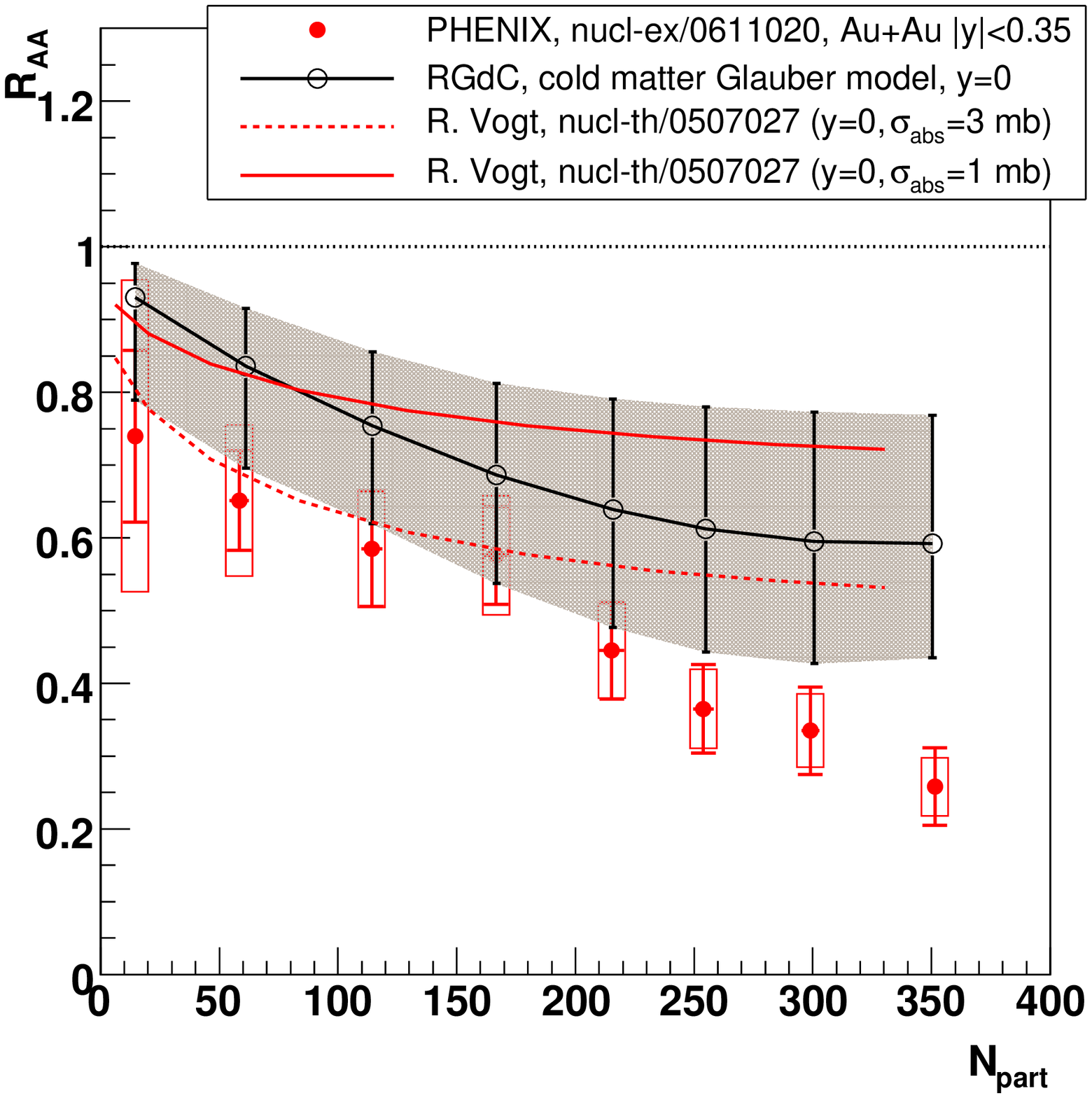}
  \end{tabular}
\end{center}
\vspace{-2ex} \caption{$R_{AuAu}$ versus
the number of participants $N_{part}$ at $y=1.7$ (left) and $y=0$ (right).
Theoretical curves are nuclear effects predictions from
Vogt~\cite{Vogt:2005ia}, solid and dashed lines being for $\sigma_{abs} = 1$ and 3~mb. The circles within the shaded bands show the prediction from the model presented
here. Data points are from PHENIX~\cite{Adare:2006ns}.
} \label{Fig2}
\vspace{-2ex}
\end{figure}

This formula assumes that a $J/\psi$ produced in a Au-Au
collision at a given rapidity $y$ suffers the product of the nuclear
effects that were observed in d-Au at this rapidity, by the ones of
the opposite rapidity
(equivalent to a Au-d collision). This assumption is at least correct for the only two effects considered so far to explain RHIC data, namely shadowing and
nuclear absorption, since quarkonia production is then proportional to the parton distribution functions ($pdf$) in each nucleus, while the average length is the sum of the length in each nucleus: $R_{AuAu} \propto pdf_1 \times pdf_2 \times \exp (-\rho \sigma (L_1+L_2))$.

The method also allows to fully propagate the errors from the d-Au measurements, both statistical and uncorrelated systematic, to the Au-Au prediction, by varying the $R_{dAu}(b)$ uncorrelated fit parameters by one standard deviation, as shown by dashed lines on figure~\ref{Fig1} left. In addition to that, two sources of systematic errors are considered. First, the fit function is not theoretically known and I tried various shapes, as displayed on figure~\ref{Fig1} right. In particular, this helps to cope up with the lack of sampling in the low impact parameter region. The obtained $R_{AuAu}$ vary by up to 5 or 12\% at forward or midrapidity, asymmetrically and depending on centrality. Second, the Glauber parameters (total cross section, Woods-Saxon parameters,...) are varied within allowed limits,which results in $R_{AuAu}$ modifications of 2 or 4\% at forward or midrapidity.

Since the approach is based on a Glauber calculation, it is easy to average
the predicted modification factors on experimental centrality classes. This is
done on figure~\ref{Fig2} where predictions are given for the Au-Au
PHENIX centrality classes from~\cite{Adare:2006ns} and compared to the measured $R_{AuAu}$.
At both rapidities, they are comparable to Vogt's predictions and incompatible with the measured $R_{AuAu}$, pointing out that $J/\psi$ are suppressed beyond cold nuclear matter effects, namely anomaly suppressed.

This method allows proper propagations of the experimental uncertainties and selections of centrality classes. Its main disadvantage lays in the fact that, to predict a given $R_{AA}$, it requires p-A (or d-A) collisions at the same energy, with the same A nucleus and for both positive and negative rapidities. In particular, these are not yet available for Cu-Cu collisions at RHIC, and not foreseen at LHC.


\section{Survival probabilities beyond cold nuclear matter}

\begin{figure}
\begin{center}
\includegraphics[width=8cm]{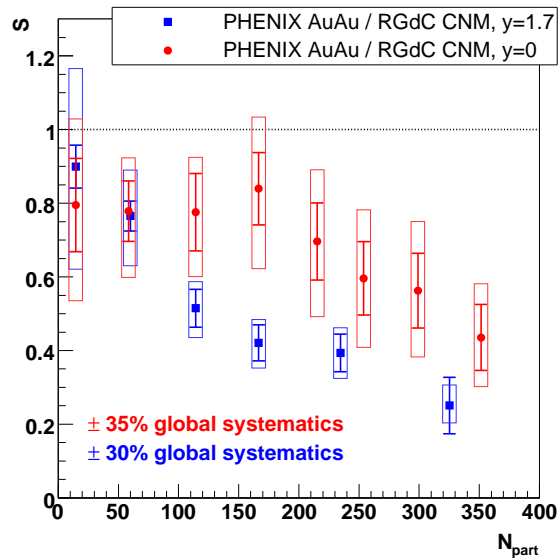}
\vspace{-2ex} \caption{$J/\psi$ survival probabilities beyond cold nuclear matter ar forward (squares, $y=1.7$) and mid (circles, $y=0$) rapidities.}
\label{Fig3}
\vspace{-2ex}
\end{center}
\end{figure}

On figure \ref{Fig3}, I divide the observed suppression by the predicted nuclear effects. The boxes stand for the Au-Au correlated errors, enhanced by the (dominant) cold nuclear matter uncertainties, while the bar is for the Au-Au statistical and uncorrelated systematic uncertainties. Additional global uncertainties of 30 or 35\% for forward or midrapidity need to be considered, mostly due to the fact that the p-p references used in $R_{dAu}$ and $R_{AuAu}$ are different and do not cancel in the ratio. A reanalysis based on  renormalized $R_{dAu}$ to the p-p reference used in $R_{AuAu}$ will substantially reduce this to $\sim$10\%.

Taking all the uncertainties into account, the most central survival probabilities are as low as $44\pm23\%$ at forward ($y=1.7$) and $25\pm12\%$ at mid ($y=0$) rapidities.

\vspace{1ex}
\indent It's my pleasure to thank Jean Gosset and Klaus Reygers for valuable help to get this work started.

\section*{References}

\bibliographystyle{myunsrt}
\bibliography{Biblio}

%
%

\end{document}